\documentclass[numberedappendix,apj,twocolumn]{emulateapj}

\usepackage[bookmarks,bookmarksnumbered,colorlinks=true, citecolor=blue, linkcolor=black,breaklinks]{hyperref}
\usepackage{epstopdf}

\newcommand{\SFRunit}{$\mathrm{M}_\odot\,\mathrm{yr}^{-1}$}

\newcommand{\newText}[1]{{#1}}

\shorttitle{Frontier Field Abell 2744 Constraints on $z\sim10$ Galaxies }
\shortauthors{Oesch et al.}

\submitted{Draft version \today}

\begin{document}

\title{First Frontier Field Constraints on the Cosmic Star-Formation Rate Density at $z\sim10$ -- the Impact of Lensing Shear on Completeness of High-Redshift Galaxy Samples
\altaffilmark{1}}

\altaffiltext{1}{Based on data obtained with the \textit{Hubble Space Telescope} operated by AURA, Inc. for NASA under contract NAS5-26555. }

\author{P. A. Oesch\altaffilmark{2},
R. J. Bouwens\altaffilmark{3},
G. D. Illingworth\altaffilmark{4}, 
M. Franx\altaffilmark{3}, 
S. M. Ammons\altaffilmark{5}, \\
P. G. van Dokkum\altaffilmark{6}, 
M. Trenti\altaffilmark{7},
I. Labb\'{e}\altaffilmark{3}
}

\altaffiltext{2}{Yale Center for Astronomy and Astrophysics, Physics Department, New Haven, CT 06520, USA; pascal.oesch@yale.edu}
\altaffiltext{3}{Leiden Observatory, Leiden University, NL-2300 RA Leiden, Netherlands}
\altaffiltext{4}{UCO/Lick Observatory, University of California, Santa Cruz, CA 95064, USA}
\altaffiltext{5}{Lawrence Livermore National Laboratory, 7000 East Avenue, Livermore, CA 94550, USA}
\altaffiltext{6}{Department of Astronomy, Yale University, New Haven, CT 06520}
\altaffiltext{7}{Institute of Astronomy and Kavli Institute for Cosmology, University of Cambridge, Madingley Road, Cambridge CB3 0HA, UK}

\begin{abstract}
We search the complete Hubble Frontier Field dataset of Abell 2744 and
its parallel field for $z\sim 10$ sources to further refine the evolution
of the cosmic star-formation rate density (SFRD) \newText{between $z\sim8$ and $z\sim10$}.
We independently confirm two images of the recently discovered
triply-imaged $z\sim9.8$ source by Zitrin et al. (2014) and set an upper
limit for similar $z\sim 10$ galaxies with red colors of $J_{125}-H_{160}>1.2$
in the parallel field of Abell 2744.  
We utilize extensive simulations to derive the effective selection volume of
Lyman-break galaxies at $z\sim10$, both in the lensed cluster field and
in the adjacent parallel field.  Particular care is taken to include
position-dependent lensing shear to accurately account for the
expected sizes and morphologies of highly-magnified sources. We show
that both source blending and shear reduce the
completeness at a given observed magnitude in the cluster,
particularly near the critical curves.  These effects have a
significant, but largely overlooked, impact on the detectability of
high-redshift sources behind clusters, and substantially reduce the expected number
of highly-magnified sources. The detections and
limits from both pointings result in a SFRD which is \newText{consistent within the uncertainties with} previous estimates at $z\sim10$ from blank fields.
The combination of these new results with all other estimates are \newText{also} consistent with a rapidly
declining SFRD in the 170 Myr from $z\sim8$ to $z\sim10$ as predicted by
cosmological simulations and dark-matter halo evolution in $\Lambda$CDM.
Once biases introduced by magnification-dependent completeness are
accounted for, the full six cluster and parallel Frontier Field
program will be an extremely powerful new dataset to probe the
evolution of the galaxy population at $z > 8$ before the advent of the
JWST.

\end{abstract}

\keywords{galaxies: evolution --- galaxies: formation ---  galaxies: high-redshift --- galaxies: gravitational lensing}

\vspace*{0.4truecm}

\section{Introduction}

The first 500 Myr after the Big Bang mark the current frontier in our exploration of cosmic history. Understanding when and how the first galaxies started to form, how they grew their stellar mass and eventually turned into the diverse population of galaxies we see today is one of the most intriguing and challenging questions of modern observational astronomy. This is the main science driver for the Director's Discretionary Time $Hubble$ Frontier Field program \citep[HFF; e.g.][]{Coe14}. The HFF will make use of lensing magnification of 4-6 foreground clusters to probe the ultra-faint galaxy population as early as 400-500 Myr after the Big Bang. Furthermore, the HFF additionally creates six deep parallel blank field pointings in order to mitigate the uncertainties of lensing magnification and cosmic variance. 

While great progress has been made recently in probing galaxy build-up out to $z\sim7-8$ \citep[e.g.][]{Bouwens11c,Bouwens14,Bradley13, Finkelstein12b,Schenker13,Mclure13,Schmidt14,BaroneNugent14}, 
beyond $z\sim8$, our understanding of galaxies is still very limited due to small number statistics.  Consequently the evolution of the cosmic star-formation rate density from $z\sim8$ to $z\sim10$ is still uncertain. The analysis of the full \newText{Hubble Ultra-Deep Field 09/12} (HUDF09/12) data and \newText{of two fields from the Cosmic Assembly Near-Infrared Deep Extragalactic Legacy Survey} (CANDELS) revealed a rapid decline of the SFRD by $\sim10\times$ in only 170 Myr from $z\sim8$ to $z\sim10$ \citep[see e.g.][but see also Ellis et al. 2013\nocite{Ellis13}]{Oesch12a,Oesch13,Oesch14}. The two detections of $z>9$ galaxies in the \newText{Cluster Lensing And Supernova survey with Hubble} \citep[CLASH;][]{Zheng12, Coe13} have not changed this broad picture of a steeper decline compared to lower redshift trends. By adding up to twelve additional very deep sightlines, the HFF program will be the prime dataset to clarify the SFRD evolution at $z>8$ before the advent of the \newText{James Webb Space Telescope \textit{(JWST)}}.

Furthermore, given the power of lensing clusters \citep[see][]{Kneib11}, the HFF program will also provide a unique dataset to study resolved morphologies of very high-redshift, multiply imaged galaxies \citep[see e.g.][]{Franx97,Kneib04,Bradley08,Zitrin11,Zitrin12,Bradley12b}, 
and will likely result in galaxy samples bright enough for  spectroscopy \citep[e.g.][]{Bradac12,Schmidt14b}. It may even be possible to probe the faint-end cutoff of the high-redshift \newText{ultra-violet} (UV) luminosity functions with the HFF dataset once all observational uncertainties and biases are under control \citep[][]{Mashian13}.

Results on $z\sim7-9$ galaxies have been reported using partial HFF data from the first observing epochs \citep[see e.g.][]{Atek14,Laporte14,Zheng14,Coe14} and very recently also from the full dataset of A2744 \citep{Ishigaki14,Atek14b}. The majority of these analyses to date have been limited, however, to the presentation of possible candidates only. The recent analysis of the complete dataset over Abell 2744 by \citet{Zitrin14} provided the first multiply imaged $z\sim10$ galaxy candidate identified from the HFF program \citep[see also][]{Ishigaki14}. The candidate JD1 is found to be a triply imaged source with an intrinsic apparent magnitude of only $\sim$29.9 mag, comparably faint as the previous $z\sim10$ galaxies identified in the deepest data over the HUDF \citep{Ellis13,Oesch14}. 
\newText{The locations of all three multiple images of JD1 are consistent with the prediction of the cluster lensing maps for a $z\sim9-10$ source, which significantly decreases the likelihood of this source being a lower redshift contaminant.}
%The fact that the magnification map predicts the location of all three images of this source that are consistent with the detections significantly increases its likelihood to be a genuine $z>9$ galaxy.

In this paper we make use of the complete HFF dataset of the first cluster, Abell 2744, and its parallel field in order to search for additional $z\sim10$ galaxy candidates and to derive the first  constraints on the star-formation rate density of $z\sim10$ galaxies based on HFF data. In particular, we will discuss the effect of shear- and position-dependent completeness for high-redshift galaxy catalogs. This proves to be very important, yet has been largely overlooked so far.
This paper is organized as follows: in Section \ref{sec:data}, we describe the dataset and sample selection. A detailed description of our completeness simulations and how shear affects the selection volume of galaxies is given in Section \ref{sec:Completeness}. Our results on the $z\sim10$ star-formation rate densities are presented in Section \ref{sec:results}, before summarizing in Section \ref{sec:summary}.

Throughout this paper, we adopt $\Omega_M=0.3, \Omega_\Lambda=0.7, H_0=70$ kms$^{-1}$Mpc$^{-1}$, i.e. $h=0.7$, consistent with the most recent measurements from Planck \citep{Planck2015}. Magnitudes are given in the AB system \citep{Oke83}, and we will refer to the HST filters F435W, F606W, F814W, F105W, F125W, F140W, F160W as $B_{435}$, $V_{606}$, $I_{814}$, $Y_{105}$, $J_{125}$, $JH_{140}$, $H_{160}$, respectively.

\section{Data and Galaxy Sample}
\label{sec:data}

\subsection{HST Dataset}

The HFF program images each cluster/blank field for 140 orbits split over seven filters with the ACS and WFC3/IR cameras. \newText{These filters are $B_{435}$, $V_{606}$, $I_{814}$, $Y_{105}$, $J_{125}$, $JH_{140}$, and $H_{160}$.} 
In this paper, we use the fully reduced version 1 HFF dataset of Abell 2744 and its parallel field provided by STScI
\footnote{\url{http://archive.stsci.edu/pub/ 
hlsp/frontier/abell2744/}}. These images were calibrated, cosmic-ray cleaned, background corrected, astrometrically aligned, and drizzled to the same output frames. In particular, we use the images drizzled at 60 mas pixel scale.

The final mosaics provided by STScI also include all ancillary data available over these fields \newText{in the same filters} from additional programs. Of particular importance is the Frontier Field UV imaging program (GO13389, PI: Siana) which adds 16 orbits of ACS data over the parallel field \newText{(split over $B_{435}$ and $V_{606}$)}. For the cluster field, we create a weighted combination of the individually provided epoch 1 and 2 ACS images using the weightmaps, which adds the pre-existing data over this cluster (GO11689, PI: Dupke).

The final $5\sigma$ depth of the images in empty regions of sky is $H_{160}=28.7$ mag as measured in circular apertures of 0\farcs4 diameter. 
For more detailed information on these data, see Koekemoer et al. (2014, in preparation) and visit the Frontier Field webpage at STScI\footnote{\url{http://www.stsci.edu/hst/campaigns/frontier-fields/}}.

Galactic extinction is accounted for by adjusting zeropoints for each HST filter using a Milky Way extinction curve \citep{Cardelli89} and $E(B-V)=0.013$ \citep{Schlegel98}. This only results in minor corrections of $<0.02$ mag in the WFC3/IR filters and up to 0.05 mag in the $B_{435}$ filter.

\subsection{Lens Models}

Gravitational lens models for all HFF clusters were produced by five teams using different methods. These are made available through the Frontier Field webpage on MAST. It is important to note that all these models are only based on ancillary data taken before the HFF campaign, and they are expected to improve and converge with the additional constraints from the many faint multiple images found in the HFF data \citep[][]{Richard14,Johnson14}.

For details on the models see, e.g., \citet{Coe14}. Here we use the five models that also released both components of the shear tensor \newText{which allow us to compute the radial and tangential magnification factors} in order to be able to properly estimate the selection volume of high redshift galaxies (see section \ref{sec:shear} \newText{for more details}). This includes the models of Bradac et al. \citep[e.g.,][]{Bradac09}, Merten et al. \citep[e.g.,][]{MertenModel}, Zitrin et al. \citep[e.g.,][]{ZitrinModel}, and Williams et al. \citep[e.g.,][]{WilliamsModel}. The results shown in the remainder of this paper are based on the lensing map provided by Zitrin et al. (Zitrin-NFW) for A2744. However, our results on the overall number densities of $z\sim10$ galaxies do not change significantly when considering other magnification maps, consistent with the findings of \citet{Coe14}.

\subsection{Removal of Intra-Cluster Light}

One significant concern with the data obtained over the cluster fields is the intra-cluster light (ICL) which significantly increases the background and limits the detectability of faint galaxies. The brightness of the ICL lies $4-5$ mag above the surface brightness limit of the HFF data over a large part of the cluster field \citep[see][]{Montes14} and thus significantly limits the direct detectability of faint sources. Furthermore, for Abell 2744 the critical curve for lensing $z\sim10$ galaxies runs partially through the ICL, which may significantly reduce the chance of finding highly magnified galaxies in standard SExtractor catalogs due to blending with the ICL and spurious detections.

In order to mitigate some of this effect, we subtract the ICL using a 2\farcs5 wide median filter. When filtering, we exclude the cores of bright sources in order to minimize over-subtraction around bright galaxies or stars. The median subtracted images are then fed to SExtractor \citep{Bertin96} to produce source catalogs using standard parameters. We found this procedure to result in somewhat more reliable catalogs and flux estimates for faint, small sources compared to running the standard SExtractor background subtraction on the original images. 
However, in future analyses, it may be possible to improve upon our treatment of the ICL using more sophisticated modeling and subtraction accounting for the ICL and bright cluster galaxies simultaneously. This may likely result in even more complete catalogs of high-redshift sources toward the cluster center. Whatever method is used, however, both the cluster galaxies themselves and the increased background due to the ICL result in reduced search volumes of high redshift galaxies in cluster images (see later Section \ref{sec:Completeness} and Figure \ref{fig:PosDepCompleteness}).

\subsection{The $z\sim10$ Lyman Break Selection}

Similar to previous selections, we identified galaxies at $z > 9.5$ by exploiting the spectral break
shortward of Ly$\alpha$ due to inter-galactic hydrogen.  Red $J_{125}-H_{160}$ colors and non-detection in shorter wavelength filters are the key features used in the selection.
In order to directly compare the HFF sample with previous analyses \citep[e.g.][]{Oesch14}, we restrict the search here to galaxies with $J_{125}-H_{160}>1.2$, which selects sources at $z\gtrsim9.5$.

We identified sources in a $\chi^2$ image constructed from the $H_{160}$ and $JH_{140}$ images and measured photometry with SExtractor run in dual image mode. 
All images were PSF-matched to the $H_{160}$ point-spread function. Colors were measured in small Kron apertures (Kron factor 1.2),
typically 0\farcs2 radius and total magnitudes were derived from larger elliptical apertures using the standard Kron
factor of 2.5, with an additional correction to total fluxes based on
the encircled flux measurements of stars in the $H_{160}$ band.

Based on these catalogs, we applied the same selections as we used previously in \citet{Oesch14}:
\begin{equation}
	(J_{125}-H_{160})>1.2
\end{equation}
\[
S/N(B_{435} \mathrm{ ~to~ } Y_{105})<2\quad \wedge \quad \chi^2_{opt+Y}\leq2.5.
\]
Furthermore, sources were required to be detected in $H_{160}$ and $JH_{140}$ with $>3.5\sigma$ in each and at least $>5\sigma$ in one of the bands.   The $\chi_{opt+Y}^2$ for each candidate source was computed following \citet{Bouwens11c} as $\chi_{opt+Y}
^2 = \Sigma_{i} \textrm{SGN}(f_{i}) (f_{i}/\sigma_{i})^2$, with $f_{i}$ the flux in band $i$ and $\sigma_i$ the associated uncertainty. SGN($f_{i}$) is equal to 1 if $f_{i}>0$ and $-1$ if $f_{i}<0$, and the
summation is over the $B_{435}$, $V_{606}$, $I_{814}$, and $Y_{105}$ bands.
The limit of $\chi^2_{opt+Y}=2.5$ efficiently excludes lower redshift contaminants while only reducing the selection volume by a small amount \citep[20\%; see also][]{Bouwens14,Oesch14}.

\begin{figure}[tbp]
	\centering	
	\includegraphics[width=\linewidth]{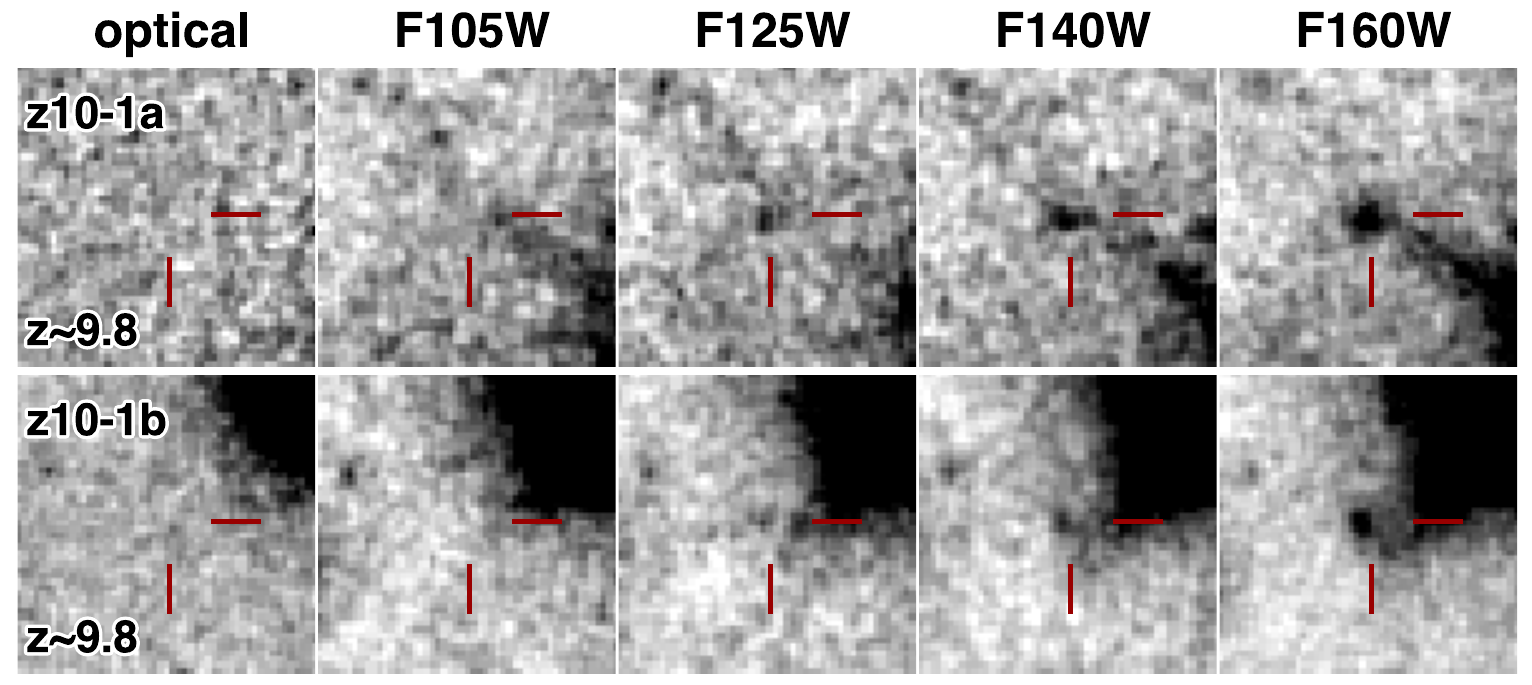}  
  \caption{\newText{3\arcsec cutouts of the two images of JD1. From left to right these show a stack of all the optical ACS bands, $Y_{105}$, $J_{125}$, $JH_{140}$, and $H_{160}$. As can be clearly seen, both source positions are affected by diffraction spikes from a nearby star and a bright cluster galaxy. To provide the most reliable estimates, we perform manual flux measurements. }.
  }
	\label{fig:stamps}
\end{figure}

\subsection{LBG Candidates in the A2744 Cluster Field}

Applying the above selection criteria to the publicly released HFF data of A2744, we identify two candidates, A2744-JD1a and A2744-JD1b. These are two images of a single, triply imaged source, independently discovered earlier by \citet[][see also Ishigaki et al. 2014]{Zitrin14}. The third image (JD1c in Zitrin et al.) lies very close to a bright foreground source and is not present in our catalogs despite aggressive deblending parameters used in our SExtractor runs. Visual inspection indicates that it is a viable source. However, it is not included in the rest of this paper, as our effective volume simulations account for sources lost due to photometric scatter or blending with neighbors, as is the case for JD1c.

The two detected images of JD1 lie on either side of the $z\sim10$ critical curve at (RA,DEC)$=$ (00:14:22.20, $-$30:24:05.3) and (00:14:22.80, $-$30:24:02.8) \newText{and are shown in Figure} \ref{fig:stamps} \citep[see also Fig 1 in][]{Zitrin14}. For both sources the photometry is heavily affected by diffraction spikes, in the first case caused by a nearby bright star and for the second source by a bright galaxy. Nevertheless, it is clear that both sources are real. While both images satisfy the color selection criteria in our standard catalog, the SExtractor photometry is likely unreliable due to the diffraction spikes. We therefore performed manual aperture photometry to confirm the color measurements and total magnitudes. In particular, in our manual measurement, we estimate the sky value in a small annulus around the source, excluding pixels which are obviously affected by the nearby diffraction spikes. Photometry is then measured in small, circular apertures of increasing size up to 0\farcs6 diameter. Finally, we use the encircled energy of stars to correct the fluxes to total as given in the WFC3 Handbook \citep{Dressel12}. 

Using this approach, we find total magnitudes of $H_{160} = 27.3\pm0.1$ and $27.2\pm0.1$ for the two images JD1a and JD1b, respectively, and colors $J_{125}-H_{160} = 1.7\pm0.5$ and $1.3\pm0.3$. These measurements are  consistent with the photometry from \citet{Zitrin14} where these images are discussed in detail and the source's photometric redshift is determined to be $z_{\mathrm{phot}}=9.8^{+0.2}_{-0.3}$. This is confirmed by the lensing geometry which satisfactorily predicts the location of the three images only if the source lies at $z>8$. 

The magnification of the two images is $\sim10\times$ as predicted by \citet{Zitrin14}. However, the full range of allowed magnification factors also predicted by other Frontier Field lensing models is $\mu\simeq4-90$. While uncertain, this source is likely to be of comparable brightness to the faintest $z\sim10$ candidates found in the XDF/HUDF12 dataset \citep{Ellis13,Illingworth13}, i.e., the $z\sim9.8$ source XDFj-38126243 \citep{Oesch13,Bouwens11a} or the $z\sim9.5$ candidate UDF12-4265-7049 \citep{Ellis13}.

\subsection{LBG Candidates in the Parallel Field}

The same search for galaxies with $J_{125}-H_{160}>1.2$ in the parallel field of the HFF cluster A2744 did not result in any candidate $z\sim10$ galaxy. While we do find a number of high-quality sources with colors within $1\sigma$ of this cut, these galaxies most likely lie at slightly lower redshift $z\sim9-9.5$ and will be discussed in a future paper. Our selection function simulations which we discuss in the next sections do statistically account for sources lost due to photometric scatter, and we therefore proceed with zero $z\sim10$ galaxy candidates from the parallel field of A2744.

\begin{figure}[tbp]
	\centering	
	\includegraphics[width=\linewidth]{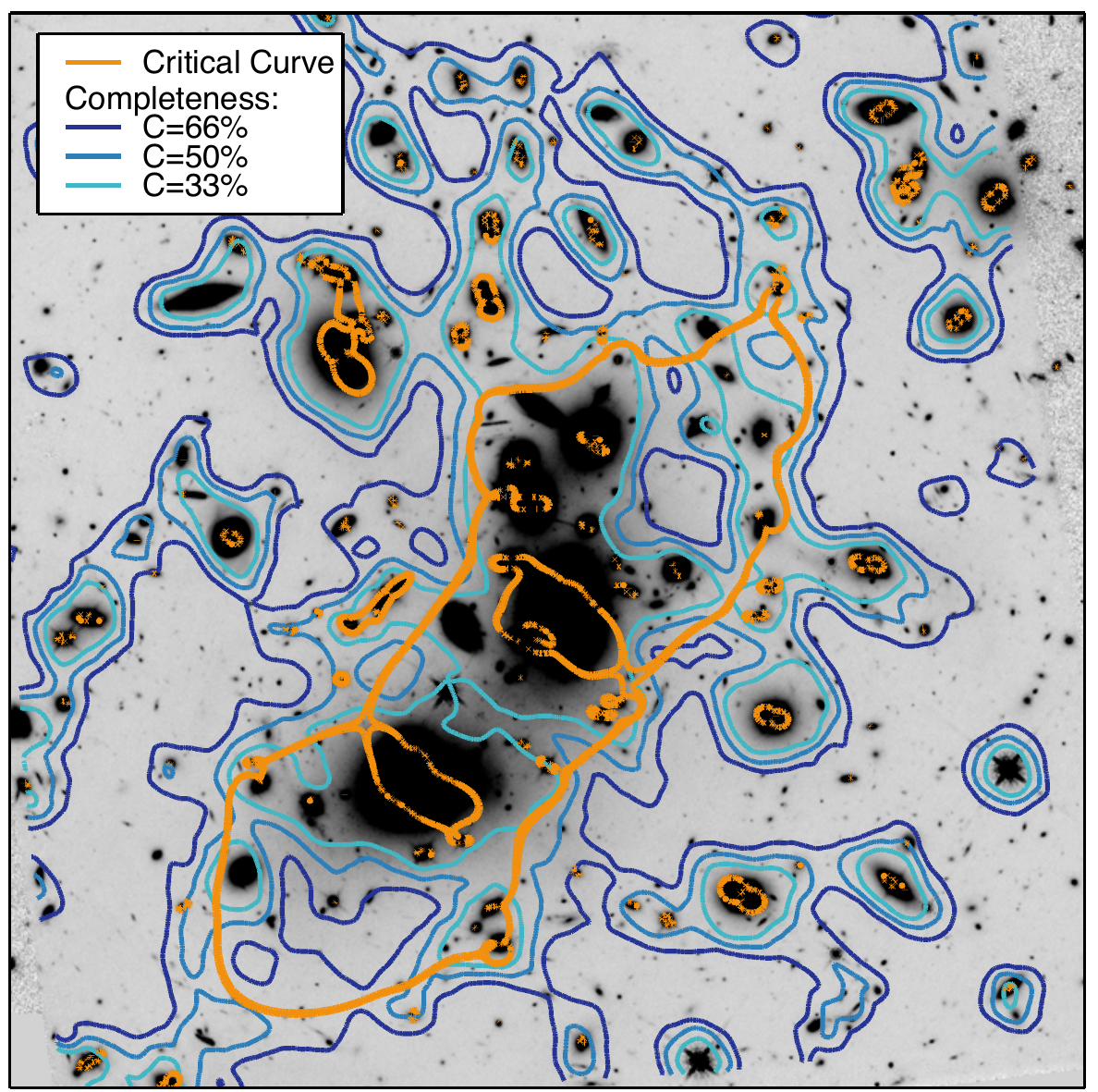}  
  \caption{Relative completeness $C$ for $z\sim10$ galaxies at fixed apparent magnitude behind Abell 2744 (blue lines). The critical curve for lensing $z\sim10$ galaxies based on the Zitrin-NFW model is shown in orange. The magnitude distribution of simulated galaxies is assumed to be flat for $H_{160}=25-28$ mag and the completeness is normalized to areas of $\mu<1.5$ (15\% of the image), where the absolute completeness is $\sim80\%$. The relative completeness over much of the cluster center is significantly reduced due to the increased background. However, lower completeness is also found around the critical curve even in the absence of bright foreground sources. This is due to the sheared morphologies of galaxies. This effect has been largely ignored in LF analyses behind lensing clusters so far. It may be possible to increase the source completeness with the use of more sophisticated modeling and subtraction of the intra-cluster light and bright foreground galaxies as well as with the adoption of a detection smoothing kernel adapted to the expected shear at a given location within the image. Nonetheless, it will not result in the same completeness levels expected in ultra-deep blank fields.
  }
	\label{fig:PosDepCompleteness}
\end{figure}

\section{Galaxy Number Densities in Lensed Fields}
\label{sec:Completeness}

In the next sections we show the importance of position-dependent source blending and shear on the completeness and selection efficiency of highly-magnified, high-redshift candidates.

\subsection{Selection Volume Accounting for Shear}
\label{sec:shear}

While faint $z\sim10$ galaxies are only marginally resolved with $HST$, they do have a finite size of $0.3-0.6$ kpc and are not point sources \citep[e.g.][]{Ono13,Holwerda14}. This has important implications for the completeness of galaxy selections around the critical curves of lensing clusters. The limited surface brightness sensitivity of HST leads to a significant reduction of the selection efficiency for the most highly magnified and sheared sources \citep[see e.g.,][]{Wong12}. 
This effect has so far been largely ignored in previous determinations of selection volumes and UV LFs behind lensing clusters due to the computational challenges involved \citep[see e.g.][]{Bradac09,Maizy10,Hall12,Coe14}, with few exceptions.
\newText{For instance, \citet{Wong12} discuss how the signal-to-noise boost from lensing depends on a galaxy's two dimensional profile due to shear. \citet{Bouwens09c} used completeness simulations which account for the mapping of galaxy images from the source plane to the image plane using the lens models. This has recently also been incorporated in new determinations of the UV LFs behind the HFF clusters \citep[e.g.][]{Ishigaki14,Atek14b}. As we show below it is crucial to include lensing shear and magnification-dependent completeness when deriving lensed number densities of galaxies.
}

We do this by taking shear into account to first order using the shear tensor to compute the tangential and radial magnification as well as the direction of the shear angle \citep[see e.g.][]{Narayan96}. We therefore rely on HFF lensing models which provide the two components $\gamma_1$ and $\gamma_2$ of the shear tensor. 
\newText{These are second order derivatives of the lensing potential $\psi$, such that $\gamma_1 = (\partial_{xx}{\psi} - \partial_{yy}{\psi})/2$, $\gamma_2 = \partial_{xy}{\psi}$ \citep[see e.g.][for a proper derivation]{Narayan96}.
}

The results in the remainder of this paper are all based on the Zitrin-NFW model. Using different models has no significant impact on the overall number densities of $z\sim10$ sources, even though the predicted magnification for individual sources can show a wide range \citep[see also][]{Coe14}. 

Based on the shear tensor, we derive the shear angle $\phi$ at each location in the image, as well as the tangential ($\mu_t$) and radial shear factors ($\mu_r$), \newText{which can be computed based on the convergence $\kappa$ and shear $\gamma$ maps provided by the lens models}:

\begin{eqnarray}
& \phi = \frac{1}{2} \arccos\frac{\gamma_1}{\gamma} = \frac{1}{2} \arcsin\frac{\gamma_2}{\gamma} &  \\
 & \mu_t = (1-\kappa-\gamma)^{-1} & \\  
& \mu_r = (1-\kappa+\gamma)^{-1} &   
\end{eqnarray}

Using these three quantities, we can estimate the effect of high magnifications on the selection function and completeness of galaxies. We follow standard procedures for blank fields and insert artificial galaxies with different light profiles, sizes, luminosities, and redshifts into the original science images. After re-running our detection algorithm with the same parameters as for the original images, the completeness $C(m)$ is simply given by the fraction of sources that are detected and observed at magnitude $m$. The only difference compared to non-cluster field completeness simulations is that we apply the position-dependent shear to the artificial galaxies before inserting these into the images. 

For computational efficiency we limit the tangential and radial shear factors to $<25$. Our estimates therefore become unreliable above $\mu\gtrsim25$ (where the completeness is overestimated; see section 3.3). This only affects a small fraction of the image plane, however \citep[$\sim$5\%; see also][]{Coe14}.

\subsection{Assumptions about the Galaxy Size Distribution}

The resulting completeness and effective selection volumes depend on the assumed properties of the simulated galaxy population. In particular, the  galaxy size and morphological profile distributions as well as the intrinsic color distributions are important parameters of such simulations \citep[see e.g.][]{Grazian11}. Here, the color distributions are set according to the luminosity dependent distribution of UV continuum slopes as measured by \nocite{Bouwens13} Bouwens et al. \citep[2013; see also][]{Bouwens12a,Rogers13,Rogers14,Finkelstein12}.

In order to account for non-regular morphologies observed in star-forming galaxies at high redshift, LBG completeness simulations often rely on scaling actual observed LBGs (e.g., at $z\sim4$) and redshifting them accounting for the difference in cosmological angular diameter distance and for intrinsic size evolution. In the case of simulations where lensing magnification is taken into account such redshifting is not possible due to the insufficient resolution of actual $z\sim4$ galaxy observations with HST to reliably reproduce higher redshift, highly magnified sources.  We therefore use another common approach adopting idealized galaxy light profiles. In practice, we use a 50\% mix of exponential disks and deVaucouleur profiles (corresponding to Sersic profiles with $n=1$ or $n=4$ respectively; Sersic 1968\nocite{Sersic68}). \newText{As shown later, this assumption has a small, but noticeable effect on the resulting completeness.}

The intrinsic size distribution of galaxies is chosen according to a log-normal distribution with mean evolving as $(1+z)^{-1}$ as is consistent with most studies of LBG size evolution at $z\sim4-10$ \citep[e.g.][]{Bouwens04a,Ferguson04,Oesch10b,Mosleh12,Ono13,Holwerda14}. 
\newText{In particular, the size distributions are normalized at $z=8$ where we set the peak of the distribution at $r=0.7$ kpc and we assume a constant width of $\sigma_{\mathrm{ln}r}=0.5$ (consistent with the distribution of halo spin parameters).}
The simulated light profiles are then sheared according to the lens model at the position where they are inserted, before convolving them with the WFC3/IR PSF.

Since we are simulating the light profiles of magnified sources, it is important to also account for any trend in size with mass or luminosity. Smaller galaxy sizes at lower luminosities are sometimes used to argue that lensing shear has no effect on galaxy completeness \citep[e.g.][]{Maizy10}. However, both the mass-size and the luminosity-size relations at high redshift are found to be very shallow following $r_\mathrm{e} \propto M^{0.17\pm0.07}$ as measured for $z\sim5$ LBGs by \citet{Mosleh12} similar to the luminosity scaling $r_\mathrm{e} \propto L^{0.25\pm0.15}$ found by \citet{Huang13}. These measurements are completely consistent with the surprisingly constant size scaling for late type galaxies at all redshifts $z=0-3$ seen in the CANDELS dataset \citep[$r_\mathrm{e} \propto M^{0.22}$; see][]{vanderWel14} and there is no convincing evidence for a change in these scaling relations at higher redshifts \citep[but see][]{Grazian12}. This suggests that a galaxy magnified by a factor $\mu=10$ is \textit{intrinsically} only $\sim1.7\times$ smaller than a non-lensed galaxy observed in the field. %We account for this in our simulations by scaling the galaxy sizes according to the magnification at the position where they are inserted in the image by $\mu^{-0.22}$. 
In order to account for this size scaling in our shear simulations over the cluster field, we scale our assumed size distribution from the blank field by $\mu^{-0.22}$ before inserting galaxies in the image. This thus corresponds to an assumed scaling of the size distribution of $r_\mathrm{e} \propto L^{0.22}$.

\begin{figure}[tbp]
	\centering	
	%\vspace{2cm}
	\includegraphics[width=\linewidth]{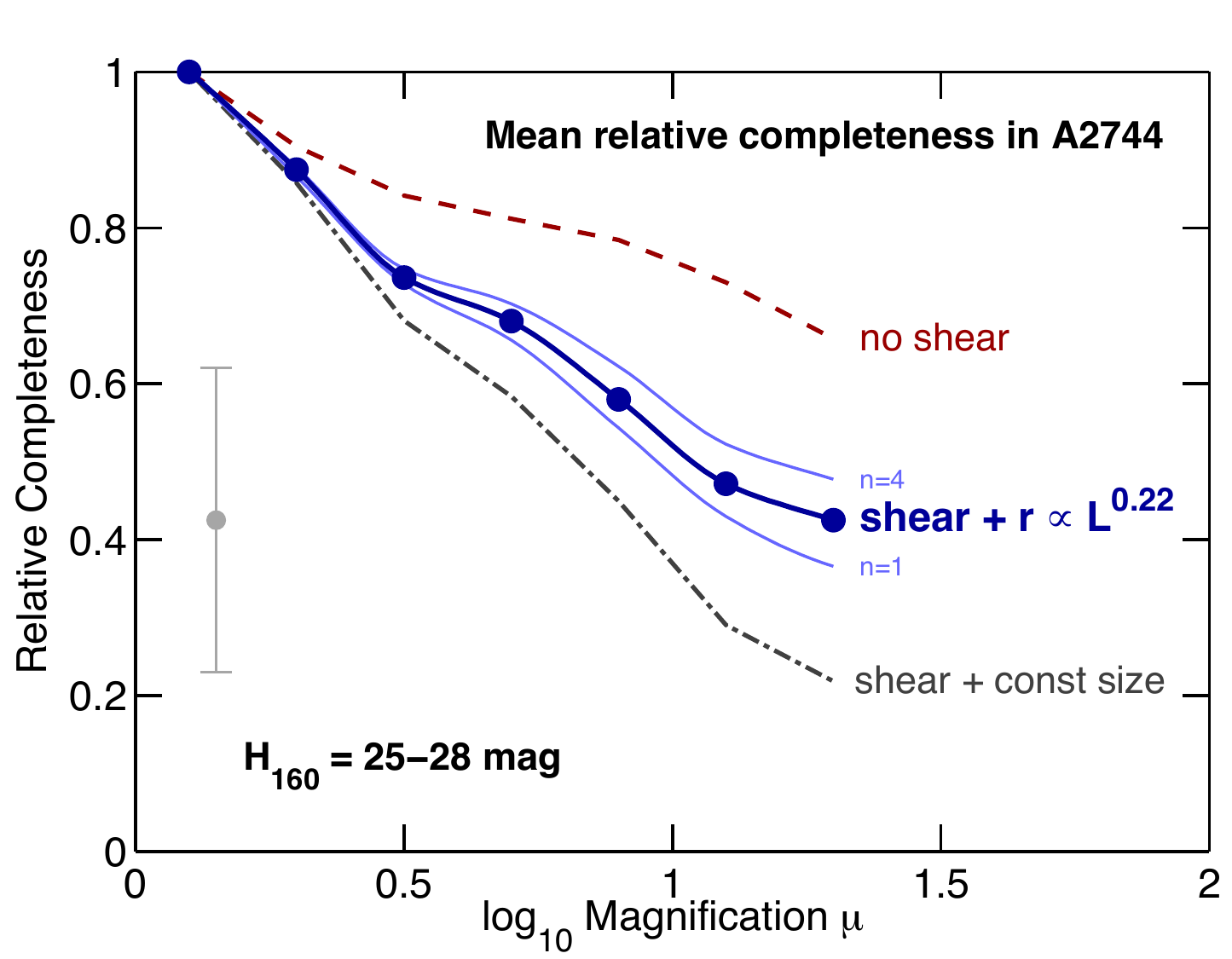}
  \caption{Mean completeness of distant galaxies in the A2744 cluster image
relative the low magnification region. The values are normalized to areas of the image of low magnification ($\mu<1.5$) corresponding to just 15\% of the area of the image in this cluster, where the completeness is about 80\%.
  The values are computed for galaxies at \textit{fixed apparent} magnitude $H_{160}=25-28$ under the assumption that galaxy sizes scale as $r_\mathrm{e}\propto L^{0.22}$ (dark blue dots and line).  Calculations are only shown up to $\mu=20$, above which our estimates start to become unreliable (overestimated) due to the use of 3\arcsec wide bins to compute the position-dependent completeness and due to our limiting the shear factors for computational efficiency. A representative gray errorbar on the left shows the 1$\sigma$ dispersion in the relation across the image.
  While lensing preserves surface brightness, highly sheared sources are spread out over many pixels resulting in a lower detection probability. The assumed size distribution has thus a significant impact on the expected completeness as shown by the gray dashed line where no size scaling with luminosity was assumed. 
Even in the absence of shear, however, magnified sources have reduced completeness due to blending with foreground galaxies and intra-cluster light as shown with the dashed red line. \newText{The faint blue lines show the impact of different galaxy profiles. The upper line shows the result using only $n=4$ Sersic profiles, while the lower curve is for exponential discs with $n=1$.}
  }
	\label{fig:MagDepCompleteness}
\end{figure}

\begin{figure}[tbp]
	\centering	
	\includegraphics[width=\linewidth]{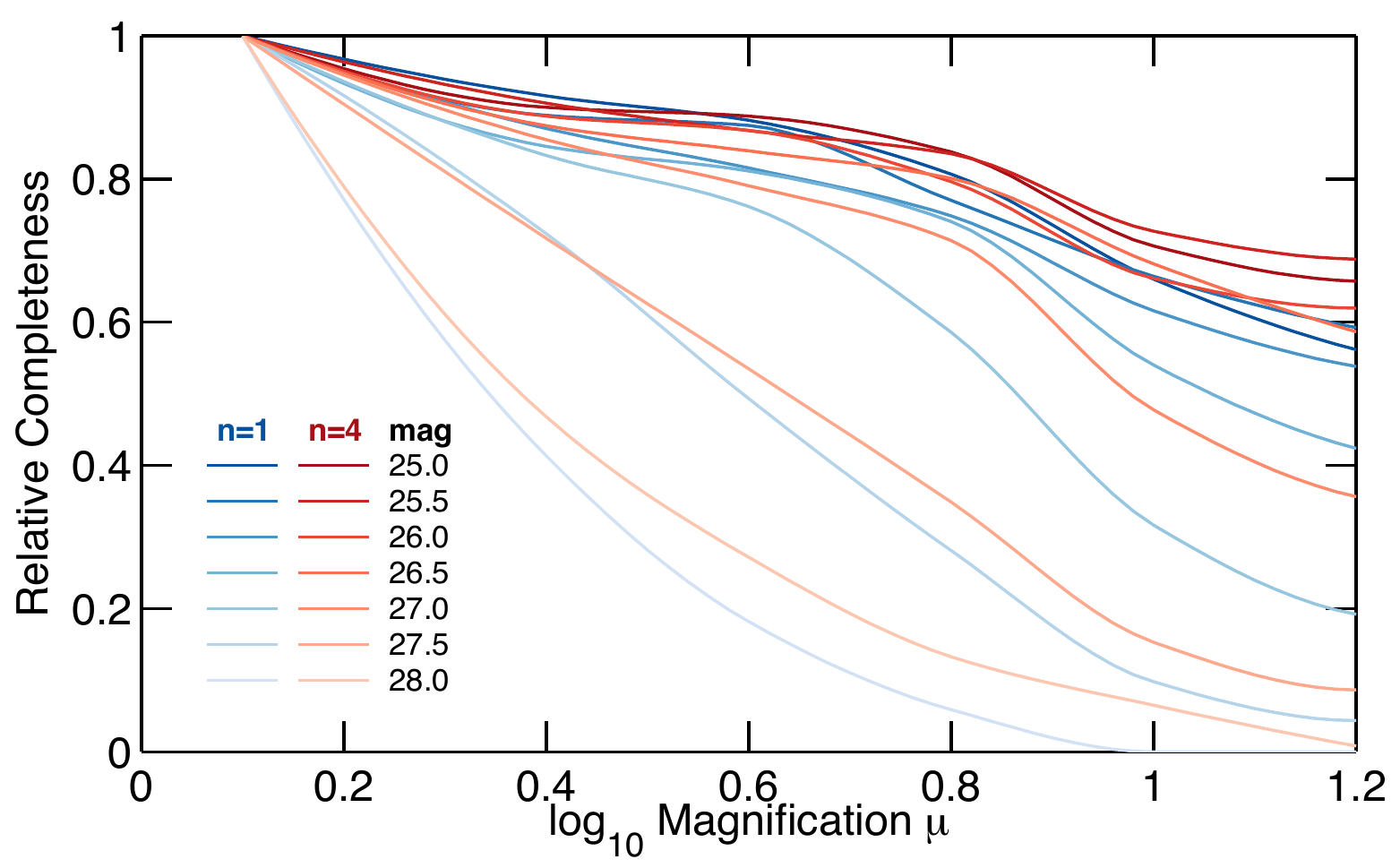}  
  \caption{\newText{Relative completeness $C$ for $z\sim10$ galaxies as a function of magnification factor $\mu$ and $H_{160}$-band magnitude behind Abell 2744. The different curves show different observed $H_{160}$ magnitude bins with decreasing luminosity (bright to faint). The colors represent different light profiles ($n=1$ blue, $n=4$ red). As seen in the previous figure, the impact of the assumption on the light profile is noticeable, but relatively small. This figure demonstrates that the reduced relative completeness for highly magnified galaxies with $H_{160}<27$ mag is mostly dominated by blending with foreground galaxies, while shear impacts sources at $H_{160}>27$ mag, i.e. about 1.5 mag from the nominal completeness limit of the data.
  }}
	\label{fig:CompletenessfMag}
\end{figure}

\subsection{Position and Magnification-Dependent Completeness}

Figure \ref{fig:PosDepCompleteness} shows the relative detection completeness for galaxies in the observed magnitude range $H_{160}=25-28$ mag as a function of position in the cluster field. The completeness is normalized to the median found in areas of the image with $\mu<1.5$ (15\% of the image), where the absolute completeness is $\sim80\%$. 

While the relative completeness decreases significantly around brighter sources in the field due to blending, it is clearly apparent that the completeness is also reduced around the critical curve of the cluster where no bright foreground sources are present. In those areas of the image, the main reason for the reduced completeness is shear and magnification.
Even though lensing conserves surface brightness, a source which is highly magnified above the survey detection limit is spread over many more pixels than a non-sheared source at the same observed magnitude, reducing its S/N and detection probability.

The relative completeness as a function of magnification averaged over the whole cluster field is shown in Figure \ref{fig:MagDepCompleteness}. As expected, we find a significant decrease in completeness toward higher magnification. 
Even though the scatter is significant in this relation, a source magnified by $\mu>10$ has on average a $\sim40-50\%$ lower chance of being detected and included in a high-redshift catalog compared to a source which is only magnified by $\mu<1.5$. 
However, as already pointed out, magnification-dependent completeness is present even when ignoring shear and magnification, simply due to blending with bright cluster galaxies closer to the critical curves (red dashed line in Fig. \ref{fig:MagDepCompleteness}). The shear adds to the incompleteness on top of this by a factor $\sim1.5\times$.

Figure \ref{fig:MagDepCompleteness} also shows that the size scaling does have a significant impact on the derived completeness relation. Using our default scaling of $r_\mathrm{e} \propto L^{0.22}$, we find that very highly magnified sources are up to a factor $\sim2\times$ more complete than assuming no size scaling at all.
The large discrepancy between these two estimates, however, shows that accurate size scaling relations are necessary to accurately compute the selection volumes of high-redshift galaxies, adding to the uncertainties in LFs estimated from cluster fields.

One possible strategy for mitigating these uncertainties is to use a differential technique to derive the relative normalization of the UV LF in various redshift bins from their relative surface densities \citep[see e.g.,][]{Bouwens12CLASH}. This is based on the assumption that the sizes and surface brightness profiles of galaxies in different redshift intervals are largely self-similar vs. luminosity. However, these assumptions in deriving the LF evolution need to be properly tested and calibrated.

\newText{Figure \ref{fig:MagDepCompleteness} also shows the impact of our assumption on the distribution of galaxy light profiles. If we simulate only Sersic $n=1$ exponential disk profiles, the resulting completeness is slightly reduced (by $\sim12\%$ at $\mu=10$) compared to the steeper $n=4$ light profile. Clearly, the impact of a size-luminosity relation is thus significantly more important, and it will be crucial to accurately determine this with future observations. }

\newText{Finally, Figure \ref{fig:CompletenessfMag} shows the magnification-dependent completeness as a function of observed magnitude and galaxy profile. This demonstrates that the drop in completeness at high magnification seen in Figure \ref{fig:MagDepCompleteness} is mostly driven by sources with $H_{160}>27.25$ mag, i.e. $\sim$1.5 mag brighter than our 5$\sigma$ detection limit. For sources with observed magnitudes brighter than this, the reduced completeness at higher magnification is mostly dominated by blending with the bright foreground cluster galaxies.}

We stress that the completeness estimates derived here only apply to galaxy catalogs using standard source detection algorithms. It may be possible to increase the source completeness around the critical curves with the use of a smoothing kernel adapted to the expected shear. Furthermore, our calculations assume idealized light profiles. Clumpy substructure in galaxies may further increase their detection probability. Quantifying these effects is beyond the scope of this paper, however.
Nevertheless, whatever detection algorithm is used, it is clear that accounting for a positional dependence of the completeness is crucial for any luminosity function or star-formation rate density analysis behind lensing clusters, which has so far been largely overlooked in the literature.

\begin{figure}[tbp]
	\centering	
	\includegraphics[width=\linewidth]{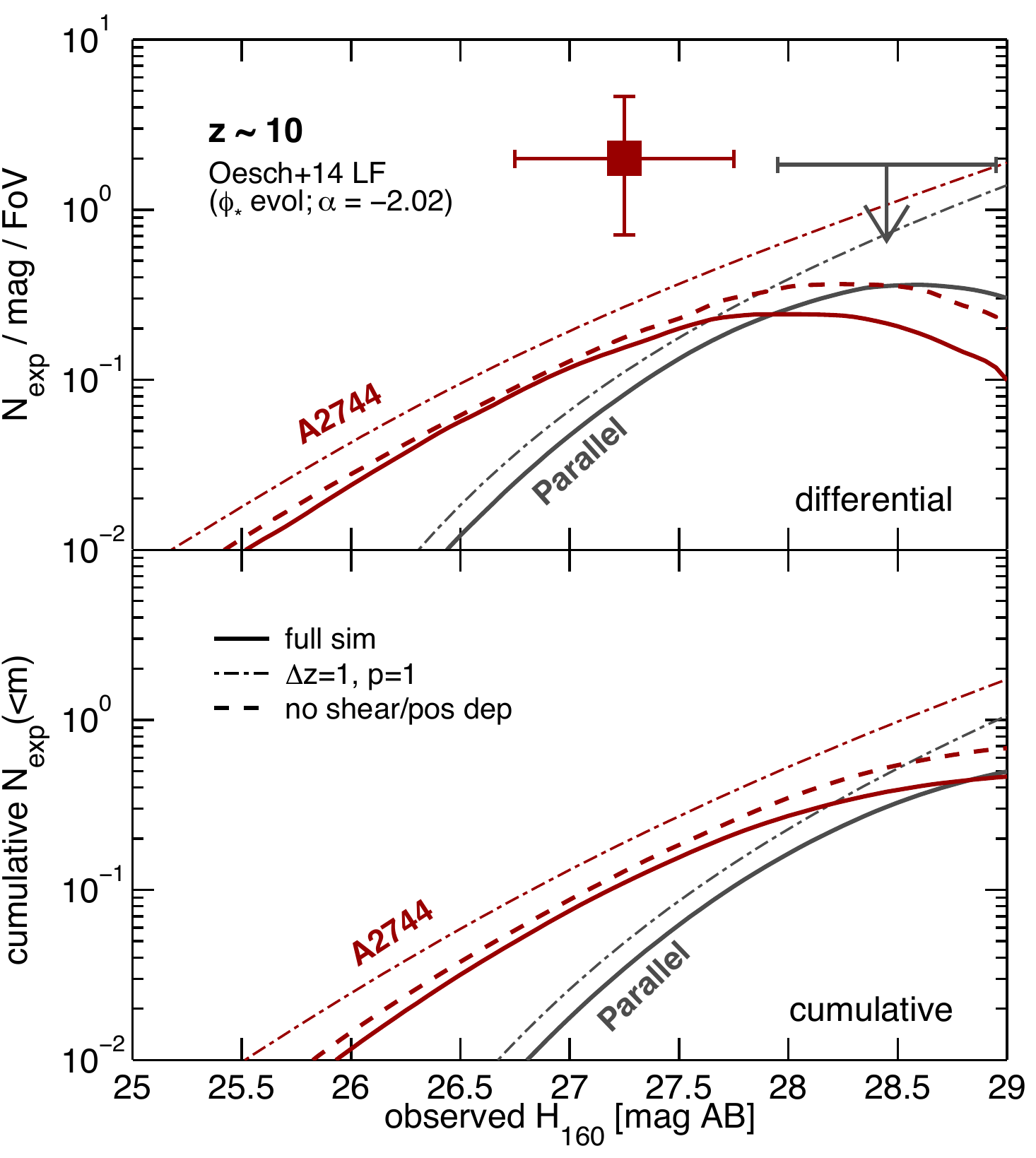}
  \caption{The number of expected $z\sim10$ candidate images in the A2744 HFF data assuming the $z\sim10$ UV LF from \citet{Oesch14}. \textit{Top} differential number counts per WFC3/IR field-of-view and per unit magnitude \textit{Bottom} cumulative number counts. Dark red lines correspond to the cluster field (assuming the magnification map Zitrin-NFW) and the dark gray lines show the parallel blank field predictions. The thick solid lines are the expected numbers based on full simulations of the selection efficiency and completeness, which also include shear for the cluster field.
  The dot-dashed lines are idealized predictions assuming a selection efficiency of $p=1$ and a selection volume of $\Delta z=1$ for comparison. The dashed line for the cluster field shows the prediction when ignoring the position-dependent completeness and assuming no shear in the simulations (as is often done in the literature). Clearly, the magnification-dependent completeness results in a significant reduction of expected images at faint magnitudes in the cluster field. The total number of expected images is comparable in both fields, in contrast to that predicted with idealized assumptions.
 The absence of $z\sim10$ galaxy candidates in the parallel field is indicated by the 1$\sigma$ upper limit in dark gray in the top panel, while the detection of two images (of the same galaxy) in the cluster field is shown as dark red square.
  }
	\label{fig:Nexpz10}
\end{figure}

\begin{figure}[tbp]
	\centering	
	\includegraphics[width=\linewidth]{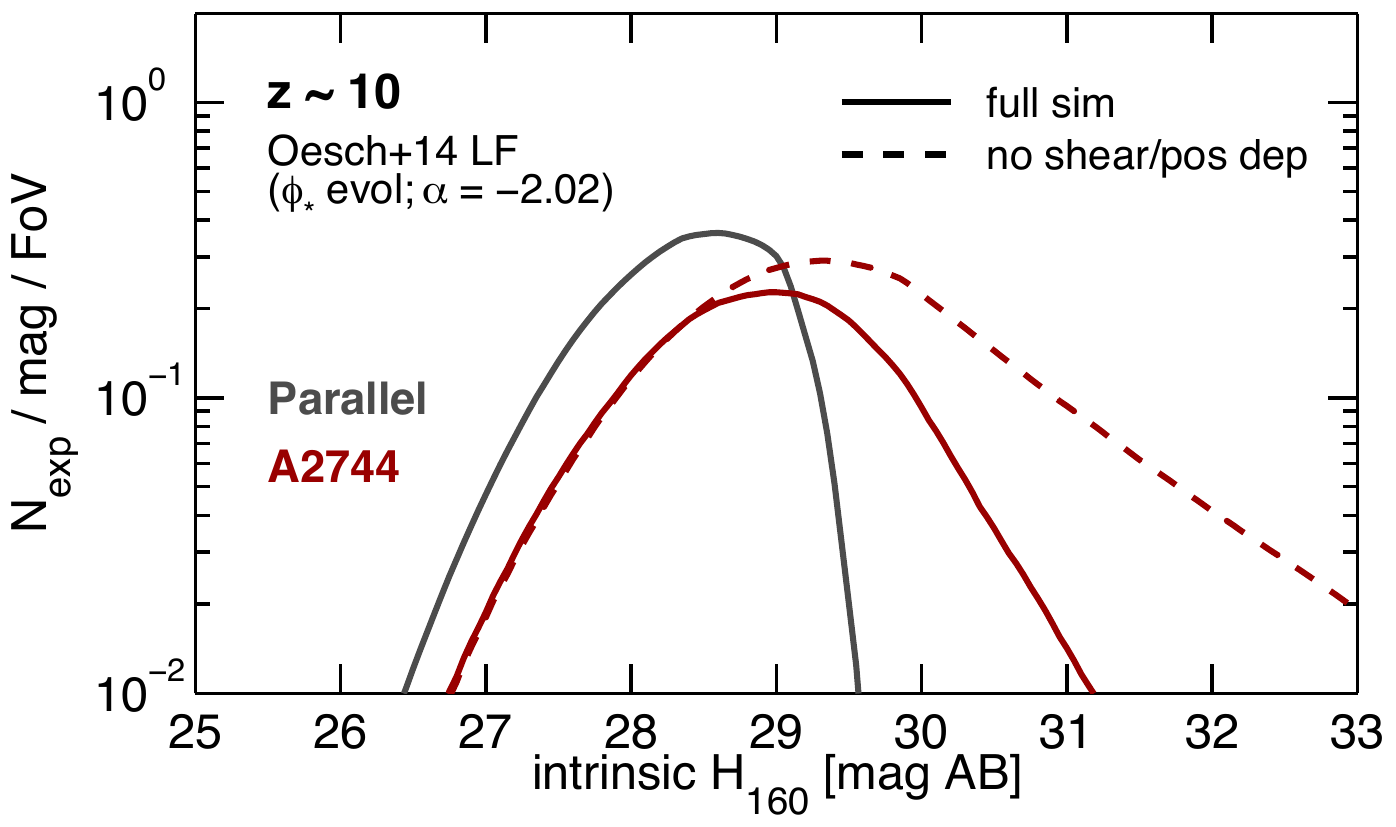}
  \caption{\newText{The differential number count of expected $z\sim10$ galaxies as a function of \textit{intrinsic}, de-lensed magnitude. As in the previous figure, dark red lines show the cluster field and dark gray lines correspond to the blank field. The thick solid lines are the expected numbers based on full simulations of the selection efficiency and completeness, which also include shear for the cluster field. By comparison to the dashed line (corresponding to a simulation without shear) it is clear that the reduced completeness due to lensing shear and foreground blending at high magnification significantly reduces the search volume at $H>30$ mag, which somewhat limits the power of clusters to probe the ultra-faint galaxy population. }
  }
	\label{fig:Nexpz10Mint}
\end{figure}

\begin{figure*}[tbp]
	\centering	
	\includegraphics[width=0.68\linewidth]{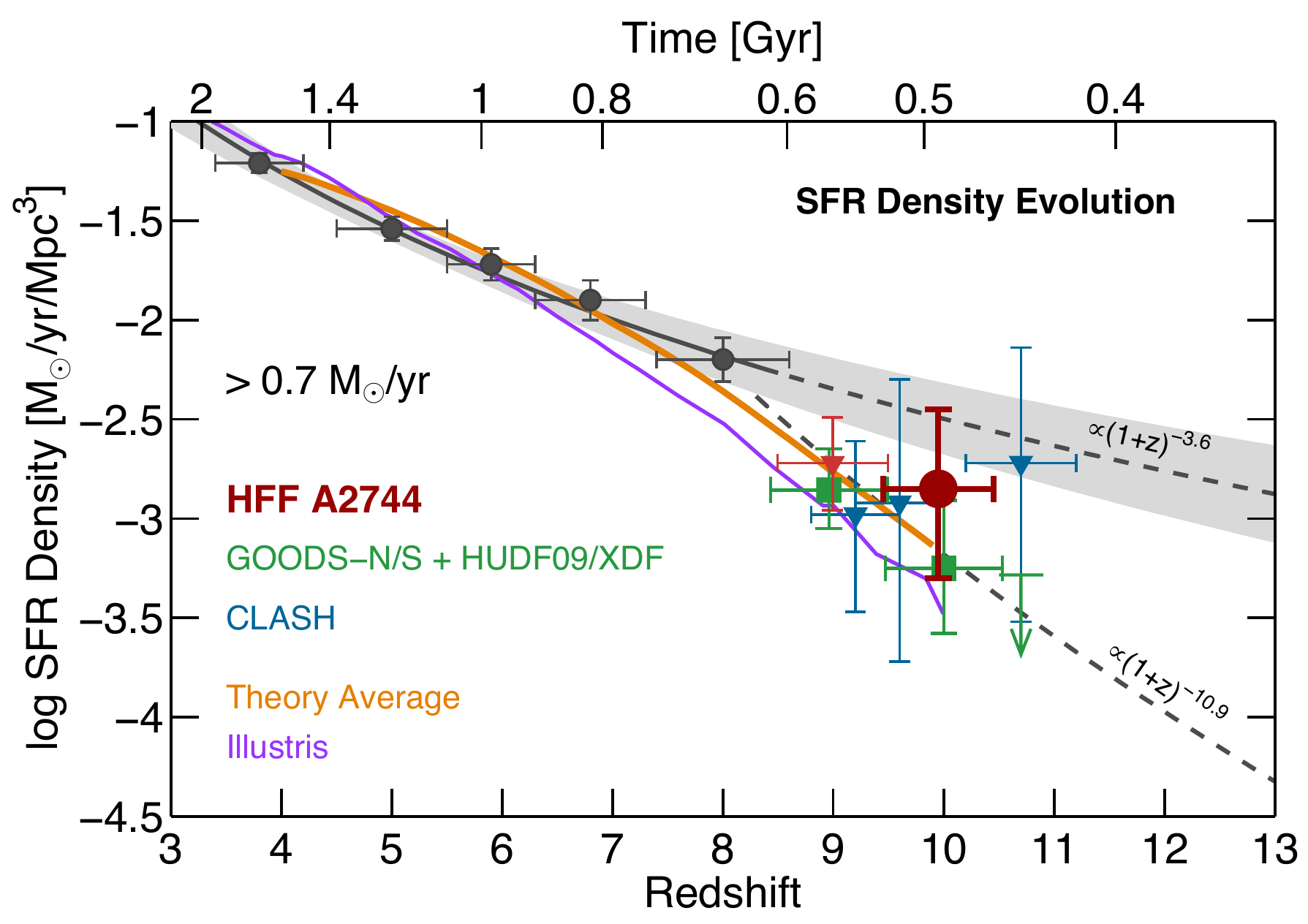}
  \caption{The evolution of the cosmic star-formation rate density at $z\sim4-11$ in galaxies down to the current detection limits in the HUDF data corresponding to $>0.7$ \SFRunit.  
  The dark red circle corresponds to the SFRD constraints from the HFF cluster A2744 and parallel field derived here, \newText{and the red triangle shows the measurement of \citep{Ishigaki14} at $z\sim9$ using A2744. The latter was computed from their $z\sim9$ LF estimate, integrated down to $M_{UV}=-17.7$.}
  Green squares show previous estimates combining the CANDELS/GOODS data with the ultra-deep imaging over the HUDF \citep[see][]{Oesch14}. Blue triangles correspond to previous estimates from CLASH cluster searches \citep{Bouwens12CLASH,Zheng12,Coe13}. 
  The lower redshift SFRD estimates are dust corrected LBG UV LFs
from \citet{Bouwens07,Bouwens12b} with $1\sigma$ uncertainty indicated by the gray band. Their empirical extrapolation is shown as the upper gray dashed line. Overall, the data are more consistent with a faster decline, as found in \citet{Oesch14}. This is indicated by the lower dashed line.
  The orange line shows an average of several theoretical model predictions shown in Figure 11 of \citet{Oesch14}. These include semi-analytical/empirical models \citep{Trenti10,Lacey11,Tacchella12} and SPH simulations \citep{Finlator11a,Jaacks13}.
   Also shown is the SFRD of the Illustris simulation \citep[purple line;][]{Vogelsberger14,Genel14}, slightly shifted to account for IMF differences in converting UV luminosities to SFRs. All these theoretical models agree with each other within $\pm0.2$ dex, and reproduce the rapid decline in the observed cosmic SFRD at $z>8$ very well. 
  }
	\label{fig:SFRDevol}
\end{figure*}

\subsection{Expected Galaxy Counts in Lensed Fields}

Clearly, the reduced completeness at high magnification also somewhat reduces the power of lensing clusters to probe deeper down the LF than ultra-deep blank fields. In this section, we estimate how this affects the expected number of $z\sim10$ galaxy candidates in the A2744 HFF data set.

Using the previous simulations we compute the magnification-dependent selection efficiency, $p(z,m,\mu)$. This is given by the fraction of inserted galaxies at magnification $\mu$ with redshift $z$ and observed magnitude $m$, which are both detected and satisfy our color selection criteria. This therefore accounts both for completeness at a given observed magnitude, as well as for photometric scatter which statistically removes galaxies from our LBG color selection box. 

Using this selection function, we can compute the number of expected galaxy images (double-counting multiple images) in bins of magnitude for a given UV LF $\phi$ from 
\begin{equation}
\frac{dN(L_\mathrm{obs})}{dm}=\int d\mu \frac{\phi(L_{obs}/\mu)}{\mu}\frac{d\Omega}{d\mu} \int dz ~ p(z,L_{obs}, \mu) \frac{dV}{dzd\Omega} 
\end{equation}
where \newText{$L_{obs}$ is the observed, magnified luminosity of a source}, $\mu$ is the magnification and $\frac{d\Omega}{d\mu}$ is the image solid angle (i.e. observed pixels) which is magnified by $\mu$.  
$dV/dzd\Omega$ is the cosmological volume per unit solid angle and redshift.
The same equation also holds for blank fields, where $\mu=1$ everywhere.

From the above equation it is clear that ignoring the position-dependent completeness and the reduction of selection efficiency due to shear in cluster fields typically results in higher expected numbers than may actually be present for a given LF. This is demonstrated in Figure \ref{fig:Nexpz10}, where we show the number of expected galaxy images for both the cluster field and the parallel blank field of A2744. 

The LF for this figure is taken from the analysis of the full CANDELS-Deep and XDF/HUDF09 dataset. This is still uncertain due to the small number of candidates and \citet{Oesch14} therefore derive two possible $z\sim10$ UV LFs based on the previous data, one in which they assume evolution in $M_*$ relative to the $z\sim8$ LF and one where only the normalization $\phi_*$ is evolving. Both derivations of this LF have an extremely steep faint-end slope $\alpha=-2.02$. For Fig. \ref{fig:Nexpz10}, we show the results for the LF which evolves only in $\phi*$, \newText{which fits the CANDELS and XDF/HUDF09 datasets better}. In addition to the curve resulting from the full selection function simulation, we also show an idealized prediction assuming $p=1$ and integrating the volume over $\Delta z=1$. This overpredicts the expected surface density distribution of candidates by a factor $\sim2\times$ for bright galaxies for both pointings.

For the cluster field, we further show the expected number of sources if shear- and magnification-dependent completeness due to blending are ignored (as is often done in the literature). Including shear reduces the total expected number of sources in the cluster field by a factor 1.6$\times$ for this assumed LF.

\newText{Figure \ref{fig:Nexpz10Mint} shows the differential number counts expected for $z\sim10$ galaxy candidates as a function of \textit{intrinsic}, unlensed magnitude. Due to the average magnification in the cluster field, the number counts peaks at about 0.5-1 magnitude fainter than in the parallel field. However, our simulations show that accounting for magnification-dependent completeness and shear significantly limits the power of lensing clusters to probe galaxies fainter than $H>30.5$ mag, i.e. intrinsically much fainter than the Hubble-Ultra Deep Field. }

Idealized calculations show that the cumulative galaxy number counts are expected to be larger behind a lensing cluster than in the field if the effective slope of the LF is steeper than $\alpha_\mathrm{eff}=-2$ in which case the reduced observed solid angle due to lensing is more than compensated for by the large abundance of faint, lensed galaxies \citep[e.g.][]{Broadhurst95}. Given our assumed UV LF with a faint-end slope of $\alpha=-2.02$ the cluster field would thus be expected to show a significantly larger number of high-redshift sources at all magnitudes. However, once we include magnification-dependent completeness, the cluster field in fact shows a very similar total number of expected images of $z\sim10$ galaxy candidates as the blank field (within $<$15\%). \newText{For the best-fit LF evolving in $\phi_*$ from \citet{Oesch14} shown in Figure \ref{fig:Nexpz10}}, we predict to detect 0.46 images in the cluster field and 0.49 $z\sim10$ sources in the parallel. \newText{For an LF evolving in $M*$ instead}, we predict 1.3 images in the cluster and 1.1 sources in the parallel field.  

If these numbers are similar for all the other five HFF clusters, the Frontier Field program is thus expected to find between 6 to 14 new $z\sim10$ galaxy candidates assuming the two different $z\sim10$ UV LFs of \citet{Oesch14} are representative. We stress that these numbers depend strongly on the exact evolution of the UV LF at $z>8$ \citep[see also][]{Coe14}. Nevertheless, at $z\sim10$ alone, the HFFs are likely to more than double the number of reliable LBG candidates known to date.

\section{The Cosmic SFRD at $z\sim10$}
\label{sec:results}

We now combine the first HFF cluster and blank field around A2744 to derive a new, independent estimate of the cosmic SFRD. From Figure \ref{fig:Nexpz10} it is clear that the two images of JD1 behind A2744 which satisfy our selection criteria will result in a higher cosmic SFRD at $z\sim10$ than we previously determined in the CANDELS-Deep and XDF/HUDF12 data.

We estimate the HFF constraint on the $z\sim10$ cosmic SFRD from the total expected number of galaxy images per WFC3/IR field relative to the earlier $z\sim10$ UV LF estimate by \citet{Oesch14}. In particular, we use their parametrization for $\phi_*$-only evolution and search for the normalization, which reproduces two predicted images in the cluster field. 

With the assumed Schechter function parameters of $\log\phi_*=-4.27$ Mpc$^{-3}$mag$^{-1}$, $M_*=-20.12$ mag, and $\alpha=-2.02$, we predict a total of only 0.46 galaxy images in the cluster field. Considering the cluster field alone, finding two images therefore requires a higher normalization, $\phi_*$, by a factor $4.4^{+5.7}_{-2.9}$ compared to the XDF/HUDF12 LF. 
Such an increase would, however, result in a total of 2.2 predicted galaxies in the HFF parallel blank field, which is marginally inconsistent with not finding any candidate with $J_{125}-H_{160}>1.2$. 

We combine the two constraints from the HFF cluster and parallel field by multiplying the Poissonian probabilities of finding 2 or 0 sources in the two fields, respectively, for a given UV LF normalization $\phi_*$. This results in a combined best-fit of $\log\phi_*=-3.9^{+0.3}_{-0.5}$ Mpc$^{-3}$mag$^{-1}$, which is completely consistent, but $0.4\pm0.4$ dex higher than found in the ultra-deep fields. 

Using this LF normalization, we estimate a cosmic SFRD from the A2744 fields.
This is done by integrating the LF to obtain the luminosity density, which we then convert to a SFRD using the standard conversion between UV luminosity and SFR \citep{Kennicutt98}. No correction for dust extinction was applied, consistent with the expectation for very little dust in these galaxies at $z>8$. 
While we adopt the widely used approach for deriving the SFR, we note that the conversion of UV luminosity to SFR assumes both an initial mass function (IMF) as well as a star-formation history. If galaxies are significantly younger than 100 Myr or if their star-formation is very bursty \citep[as predicted by some models; e.g.][]{Dayal13,Wyithe14}, the standard conversion factor may need to be corrected \citep[see e.g.][]{Reddy12}. The investigation of these alternatives goes beyond the scope of this paper, however.

When integrating the UV LF down to $M_{UV}=-17.7$, which corresponds to a SFR limit of 0.7 \SFRunit, we obtain a SFRD of 
 $\log\dot{\rho_*}=-2.8^{+0.3}_{-0.5}$ M$_\odot$\,yr$^{-1}$\,Mpc$^{-3}$. This is shown in Figure \ref{fig:SFRDevol}, where we also plot the previous estimates for comparison. 

\newText{Our approach to estimate the SFRD, which is an integrated quantity, is not sensitive to the exact magnification of the candidate source. I.e. even if the magnification at the exact source position presented in \citet{Zitrin14} were underestimated and the galaxy intrinsically had a SFR of less than our limit of 0.7 \SFRunit, the SFRD value derived here is still valid. As long as the magnification map produces an accurate differential number count distribution (which is marginalized over the image plane) our approach of seeking the best-fitting normalization to the UV LF and integrating this to a fixed SFR limit produces accurate results. }

While the new constraint from the A2744 HFF fields is \newText{somewhat} higher than the previous ultra-deep field constraints, it is consistent with the rapid decline across $z\sim8$ to $z\sim10$ that is predicted by theoretical models. In particular Fig \ref{fig:SFRDevol} also shows the average SFRD evolution of a series of semi-analytical/empirical models \citep{Trenti10,Lacey11,Tacchella12} and from SPH simulations \citep{Finlator11a,Jaacks13} as well as the SFRD from the Illustris simulation \citep{Vogelsberger14,Genel14}. Where necessary, we shifted the theoretical models to account for our use of a Salpeter IMF when converting the UV luminosity to SFR. All theoretical models agree on a very rapid decline in the cosmic SFRD by $\gtrsim5\times$ from $z\sim8$ to $z\sim10$ when limited at $>0.7$ \SFRunit, indicating that a rapid build-up of galaxies above this limit is a generic prediction of any model of galaxy formation \citep[see also][]{Oesch14}.

Nevertheless, given the still large errorbars, a more gradual decline in the SFRD as empirically estimated based on the UV LF evolution across $z\sim4-8$ (see gray line in Fig \ref{fig:SFRDevol}) can still not be completely ruled out. If the faint-end slope of the UV LF does not steepen further at higher redshift \citep{Bouwens12b}  or if the escape fraction stays constant, this more gradual decline may be necessary for galaxies to complete reionization in agreement with the high optical depth measurement by WMAP \citep[e.g.][]{Robertson13,Kuhlen12}. Note, however, that the rapid decline in the observed SFRD may simply be a consequence of our fixed detection limit in luminosity and is likely still compatible with a more gradual evolution of the \textit{total} SFRD. This is supported both by the higher SFRD estimates of gamma ray burst counts \citep[e.g.][]{Kistler09,Trenti12b,Robertson12}, which are sensitive to the total SFRD, and by simulations \citep[e.g., compare with][]{Vogelsberger14}.

\section{Summary}
\label{sec:summary}

This paper presented a first estimate of the cosmic SFRD at $z\sim10$ based on Frontier Field data. In particular, we show that extensive completeness simulations including source blending and lensing shear close to the critical curve are crucial for any analysis of cluster data. We find a significantly lower completeness at higher magnification than for comparable blank field searches at a fixed observed magnitude (Figures \ref{fig:PosDepCompleteness} and \ref{fig:MagDepCompleteness}). This can be ascribed to several effects: blending with bright foreground cluster galaxies, higher background due to intra-cluster light, but also due to shear spreading out highly magnified sources over many pixels.

Sources at high magnification are on average only 70\% complete in the A2744 image compared to a blank field even when the effect of shear is ignored (due to blending with foreground sources and the ICL). Shear further reduces the completeness at $\mu>10$ by $\sim1.5\times$. However, the exact completeness at high magnification sensitively depends on the assumed size distribution for very faint sources below the detection limit of current blank field data (see Fig \ref{fig:MagDepCompleteness}). This effect therefore adds to the overall uncertainty of LF and SFRD estimates from cluster lensing fields.

This position-dependent completeness has \newText{often been} overlooked in the literature \citep[but see][and recently Ishigaki et al. 2014, Atek et al. 2014a]{Wong12}. However, it has important consequences on the expected number of high-redshift candidates seen behind lensing clusters compared to blank fields. 
In Figure \ref{fig:Nexpz10}, we show that the reduced completeness results in a similar number of source images predicted for the A2744 cluster and parallel field, very different from what is commonly assumed. 

Following previous blank field studies, we search the HFF A2744 cluster and parallel field data for $z\sim10$ galaxy candidates using a criterion $J_{125}-H_{160}>1.2$ and non-detections at shorter wavelength. While no candidates are found in the parallel field, we find two images of the same source lensed by the cluster \citep[previously identified in][]{Zitrin14} which both satisfy our selection criteria.

Combining the one multiply imaged candidate over the cluster field with the null detection in the parallel field, we derive a cosmic SFRD at $z\sim10$ which is consistent, but $0.4\pm0.4$ dex higher than found earlier in the ultra-deep blank fields (see Figure \ref{fig:SFRDevol}). Not surprisingly, this independent measurement based on the first completed HFF cluster does not allow us to significantly rule out different possible scenarios for the SFRD evolution between $z\sim8$ and $z\sim10$. 
The combination of these new results with all other estimates from the literature remain consistent with a rapidly
declining SFRD as is predicted by cosmological simulations and dark-matter halo evolution in $\Lambda$CDM.

The completed multi-cluster HFF dataset will allow to further increase the sample size of galaxies at $z\sim10$ and to significantly tighten this first estimate of the cosmic SFRD.
Once biases due to magnification-dependent incompleteness are taken into account, the HFF survey will be a key dataset to study the galaxy population at $z>8$ before the advent of the $JWST$.

\vspace*{0.5cm}

\acknowledgments{ We thank the anonymous referee for very helpful feedback and suggestions, which greatly improved this paper.
This work was supported by NASA grant NAG5-7697 and NASA grant HST-GO-11563.01. 
We are grateful to the directors of STScI and SSC to execute the dedicated Frontier Field program. 
We thank Shy Genel and Mark Vogelsberger for providing and discussing the results of the Illustris simulation. 
This work utilizes gravitational lensing models produced by PIs Bradac, Ebeling, Merten \& Zitrin, Sharon, and Williams funded as part of the HST Frontier Fields program conducted by STScI. STScI is operated by the Association of Universities for Research in Astronomy, Inc. under NASA contract NAS 5-26555. The lens models and datasets were obtained from the Mikulski Archive for Space Telescopes (MAST).}

Facilities: \facility{HST (ACS, WFC3)}.

%\bibliography{MasterBiblio}
%\bibliographystyle{apj}

\end{document}